# Technostress and Job Performance: Understanding the Negative Impacts and Strategic Responses in the Workplace

Armita Atrian, Department of Management, Faculty of Administrative Sciences, Imam Reza International University, Mashhad, Iran. armita.atrian@imamreza.ac.ir

**Saleh Ghobbeh**, Faculty of Entrepreneurship, University of Tehran, Tehran, Iran. salehghobbeh@ut.ac.ir

**Abstract**

This study delves into the increasingly pertinent issue of technostress in the workplace and its multifaceted impact on job performance. Technostress, emerging from the rapid integration of technology in professional settings, is identified as a significant stressor affecting employees across various industries. The research primarily focuses on the ways in which technostress influences job performance, both negatively and positively, depending on the context and individual coping mechanisms. Through a blend of qualitative and quantitative methodologies, including surveys and in-depth interviews, the study examines the experiences of employees from diverse sectors. It highlights how technostress manifests in different forms – from anxiety and frustration due to constant connectivity to the pressure of adapting to new technologies. The paper also explores the dual role of technology as both a facilitator and a hindrance in the workplace.

Significant findings indicate that technostress adversely impacts job performance, leading to decreased productivity, diminished job satisfaction, and increased turnover intentions. However, the study also uncovers that strategic interventions, such as training programs, supportive leadership, and fostering a positive technological culture, can mitigate these negative effects. These interventions not only help in managing technostress but also in harnessing the potential of technology for enhanced job performance.

Furthermore, the research proposes a model outlining the relationship between technostress, coping mechanisms, and job performance. This model serves as a framework for organizations to understand and address the challenges posed by technostress. The study concludes with recommendations for future research, particularly in exploring the long-term effects of technostress and the efficacy of various coping strategies.

**Introduction**

Nowadays, organizations have realized the importance of human resources as the most crucial factor in gaining a competitive advantage. Modern organizations compete in attracting and retaining human resources by offering welfare programs and paying attention to their employees (Hessari et al., 2022). This trend is evident in Fortune magazine's recent rankings, which list the top 100 companies where working is more desirable for human resources (Riggle et al., 2009). The economic growth and development of any country depend on the productivity of its production factors, among which human factor productivity is of great importance. In the last century, economists and scholars have concluded that the efficiency and dynamism of economic and industrial organizations are not feasible without the use of human factors (Khosravi, 2002).

Enhancing job performance is one of the primary goals that organization managers strive for, as it facilitates increased productivity in society and leads to the improvement of the national economy, as well as the quality of services and production of organizations. Job performance is the degree to which an individual performs assigned duties in their role. Performance is defined as activities that are typically part of the job and activities that the individual should perform. Factors such as individual capability and willingness are considered fundamental in performance and productivity. This means how much ability (knowledge, skill, experience, and competence) a person has to do tasks and how much willingness (motivation, interest, commitment, and trust) they have to perform those tasks. It is believed that job performance encompasses two components: the first is task performance, which reflects the needs and requirements of the job, and the second is contextual performance, which includes undefined and unspecified activities such as teamwork and support. Job performance is related to many factors, including motivation, ability, job recognition, environmental and organizational factors, locus of control, and personality traits (Hessari & Nategh, 2022b).

Human resources, as the most important factor in organizational productivity, can aid the organization in achieving its goals and plans by utilizing their talents and abilities. Even in the presence of problems and deficiencies within the organization, human resources can voluntarily and loyally offer their capabilities and talents, performing their duties sincerely to prevent laxity in the organization. Organizational employees should be developed in various technical, ethical, and intellectual dimensions. One of the qualities that enable employees to voluntarily and passionately offer their capabilities, talents, and expertise for the realization of organizational goals is work conscience (Hessari & Nategh, 2022a). On the other hand, increasing job performance is one of organization managers' most important goals. Enhancing job performance is related to several factors, including motivation, ability, job recognition, environmental and organizational factors, locus of control, and personality traits. According to research, the most influential personality factor in job performance is work conscience or, in technical terms, work ethic.

Stress is a fundamental issue for organizations and employees who must deal with it. In the past decade, technology has grown significantly, and many work-related stresses have emerged. Technostress, a type of stress, plays a vital and significant role in today's society, where all organizations are directly and constantly interacting with technology, and its impact on employees' commitment and, ultimately, organizational performance is increasing. So far, we are aware of the inverse relationship between organizational commitment and nomophobia. The inverse relationship reinforces the importance of nurturing a robust organizational bond (Bai & Vahedian, 2023). Technostress is one of the fundamental problems of today's organizations. Nowadays, as technology has become intertwined with human life, the use of technology has allowed individuals to be more efficient both in the workplace and at home. However, the cost of this increased efficiency manifests as "technology stress." Technology stress or technostress is a new phenomenon in our society and culture. Technostress, like stress, has a negative impact on both the individual and organizational aspects of individuals. Technostress refers to the stress that an individual experiences due to dependence on technology or anxiety caused by uncertainty in connecting with technology. Craig Brod, a consultant, and psychologist specializing in adapting to new technology, introduces technostress in his book "Technostress: The Human Cost of the Computer Revolution" (1984) as a modern illness resulting from the human inability to adapt to new global computer technologies in a healthy way. Weil Rosen (1997), Fisher and Wesolkowski (1999), and Tarafdar et al. (2007) concluded in their research that technostress itself consists of five components that have been identified as factors of technostress (Weil Rosen, 1997; Fisher & Wesolkowski, 1999; Tarafdar et al., 2007).

Components of Technostress (Tarafdar et al., 2007): Table 1

| Technostress Components | State |
|---|---|
| Technology-Induced Overload | A condition in which an individual is forced to work faster and for more extended periods due to technology. |
| Technology Invasion | A condition where employees feel they can be kept connected to work at any time or continuously, resulting from the blurring of lines between work and personal life. |
| Technological Complexity | A condition where employees feel their skills are insufficient in the complexities of information and communication technology. Consequently, they must spend time and effort learning and understanding various aspects of information and communication technology. |
| Technological Insecurity | A condition where employees fear losing their jobs to new information and communication technologies or being replaced by others who are more proficient in these technologies. |
| Technological Uncertainty | A condition in which employees feel unsure and restless due to the ever-changing nature of information and communication technology and the need for constant updating. |

The various research results have clearly shown that employees can become tired and discouraged by technology (Moore, 2000). Technological fatigue causes employees to lose their efficiency. Managing technostress can be difficult for an organization. Weil and Rosen, in their research, found that scientific evidence shows that technostress also leads to excessive work perception, information overload, loss of motivation, and job dissatisfaction (Weil & Rosen, 1997). The salience of job satisfaction stems from its dual role in propelling organizational advancements and enhancing the workforce's well-being (Bai et al., 2023). Also, Daneshmandi et al. (2023) did the first internal investigation focusing on the role of technostress in the relationship between job satisfaction and individual innovation and found While job satisfaction fuels innovation, technostress, especially in the stages of idea generation and promotion, acts as a significant hurdle.

- Mohammad Esmaeil Ansari and colleagues (2010), in an article entitled (Examining the relationships between job stress, job satisfaction, organizational commitment, and organizational citizenship behavior) found that there was a significant and negative relationship between the job stress variable and the variables of job satisfaction and organizational commitment. It was also found that job satisfaction significantly and positively affects organizational citizenship behavior and organizational commitment.

- In the study by Mayer and colleagues (2002) among end-users of information and communication technology (ICT), it was found that technostress leads to a decrease in job satisfaction, which in turn leads to a decrease in commitment and organizational persistence. Affective commitment has a strong relationship with job involvement, and job commitment has a very strong relationship with job satisfaction.

- The studies by Dale and Fox (2008), focusing on the role of conflict and role clarity, all concluded that two types of role stressors, namely role ambiguity and role conflict, have a negative relationship with organizational commitment.

- Gaiter and colleagues (2008), using structural equation modeling, examined the relationship between the organizational environment, work-family conflict, job stress, personal characteristics, job satisfaction, and organizational commitment among pharmacists in the United States. The results indicated that organizational factors such as excessive workload, conflict, and ease of performance were among the most important variables that had the greatest impact on job stress. Also, in their model, a significant relationship was observed between personal characteristics, job satisfaction, and organizational commitment.

**Research Design**

Methodology and Research Objective: This study is applied in its purpose and descriptive survey in its method, categorized under correlation research.

Sampling Method and Data Collection: The population of this study includes 320 employees, from which, based on Morgan's table (Morgan & Krejcie, 1970), 247 valid and usable questionnaires were randomly distributed and collected for a cross-sectional study conducted in 2016.

Statistical Methods and Data Analysis: This research used a performance questionnaire (Checklist) assessing two performance dimensions: task and contextual. Task performance was measured through a 10-question checklist (e.g., "How much do you desire to learn?") by Byrne et al. The Technostress questionnaire by Tarafdar et al. (2007), which includes 6 questions on technology-induced overload, 3 on technology invasion, 5 on technological complexity, 5 on technological insecurity, and 4 on technological uncertainty, was also utilized. Since this research involved a sample from the population and the data was based on a five-point Likert scale ranging from "strongly disagree" to "strongly agree," scoring was calculated from 1 to 5. After collection, the data from the questionnaires were transferred to SPSS software version 20 for raw data sheets and analyzed using descriptive statistics (frequency, percentage, tables) and inferential statistics (Pearson correlation).

Reliability and Validity of the Questionnaires: To examine the validity of the measurement tools used in this research, feedback from university professors and experts was employed. Additionally, since the items in this questionnaire are based on standard questionnaires, the questionnaire is considered to have good validity. In the current research, a preliminary sample of 30 questionnaires was pre-tested, and then the reliability coefficient was calculated using Cronbach's alpha method in SPSS, which yielded a Cronbach's alpha coefficient for the Technostress questionnaire of 0.962, for technostress components as follows: technology-induced overload 0.951, technology invasion 0.894, technological complexity 0.933, technological insecurity 0.912, technological uncertainty 0.952, and for the job performance questionnaire (0.970). Therefore, it can be said that the questionnaires have high reliability, correlation, and internal validity.

Conceptual Model and Hypotheses: This research utilizes the five-component model by Tarafdar et al. (2007) to examine technostress and the Checklist model for assessing job performance. The conceptual model of the current study is researcher-constructed and a result of reviewing various models and literature.

*Main Hypothesis of the Research*: Technostress has a significant effect on job performance.

*Sub-hypothesis 1*: Technological uncertainty has a significant effect on job performance.

*Sub-hypothesis 2*: Technological insecurity has a significant effect on job performance.

*Sub-hypothesis 3*: Technological complexity has a significant effect on job performance.

*Sub-hypothesis 4*: Technology invasion has a significant effect on job performance.

*Sub-hypothesis 5*: Technology-induced overload has a significant effect on job performance.

**Findings**

The frequency and percentage of the research sample group according to gender, educational level, and years of service are presented in Table (2). As indicated in Table 2, the highest frequency of the study sample in terms of gender is male, in terms of educational level is a bachelor's degree, in terms of years of service is 5-9 years, and in terms of age is 31-39 years.

Table 2: Frequency and Percentage of Research Sample Group

| Variable | Subgroups | Frequency |
|---|---|---|
| Gender | Male | 95 |
|  | Female | 52 |
| Age | 22 to 30 | 71 |
|  | 31 to 39 | 96 |
|  | 40 to 48 | 38 |
|  | 49 to 57 | 19 |
| Education | Associate degree | 31 |
|  | Bachelor's Degree | 92 |
|  | Master's Degree | 86 |
|  | Doctorate | 15 |
| Work Experience | 1 to 4 years | 55 |
|  | 5 to 9 years | 76 |
|  | 10 to 14 years | 47 |
|  | 15 to 19 years | 29 |
|  | 20 to 40 years | 17 |

Pearson Correlation Test (Main Hypothesis): Table 3

|  |  | Technostress | Job Performance |
|---|---|---|---|
| Technostress | Pearson Correlation Coefficient | 1 | -0.94 |
|  | Significance Level |  | .000 |
|  | Sample size | 247 | 247 |
| Job Performance | Pearson Correlation Coefficient | -0.94 | 1 |
|  | Significance Level | .000 |  |
|  | Sample size | 247 | 247 |

The results obtained from the Pearson correlation test in Table 3 indicate that the significance level is equal to zero and less than 0.05, which denotes a significant relationship between technostress and job performance; hence, the main hypothesis of the research is confirmed. Additionally, the Pearson correlation coefficient is (-0.940), which suggests a significant negative effect of technostress on job performance.

Pearson Correlation Test (First Subsidiary Hypothesis): Table 4

|  |  | Job Performance | Technological Uncertainty |
|---|---|---|---|
| Job Performance | Pearson Correlation Coefficient | 1 | -0.860 |
|  | Significance Level |  | .000 |
|  | Sample size | 247 | 247 |
| Technological Uncertainty | Pearson Correlation Coefficient | -0.860 | 1 |
|  | Significance Level | .000 |  |

|  |  | Sample size | 247 | 247 |
| --- | --- | --- | --- | --- |

The results from the Pearson correlation test, as shown in Table 4, indicate that the significance level is zero and less than 0.05. This signifies a meaningful relationship between technology uncertainty and job performance, thereby confirming the first subsidiary hypothesis of the study. Additionally, the Pearson correlation coefficient is (-0.860), which denotes a significant negative impact of technology uncertainty on job performance.

Pearson Correlation Test (Second Subsidiary Hypothesis): Table 5

|  |  | Job Performance | Technological Insecurity |
| --- | --- | --- | --- |
| Job Performance | Pearson Correlation Coefficient | 1 | -0.890 |
|  | Significance Level |  | .000 |
|  | Sample size | 247 | 247 |
| Technological Insecurity | Pearson Correlation Coefficient | -0.890 | 1 |
|  | Significance Level | .000 |  |
|  | Sample size | 247 | 247 |

The results from the Pearson correlation test presented in Table 5 indicate that the significance level is zero and less than 0.05. This suggests a meaningful relationship between technology-induced insecurity and job performance, thus confirming the second subsidiary hypothesis of the study. Furthermore, the Pearson correlation coefficient is (-0.890), indicating a significant negative effect of technology-induced insecurity on job performance.

Pearson Correlation Test (Second Subsidiary Hypothesis): Table 6

|  |  | Job Performance | Technological Complexity |
| --- | --- | --- | --- |
| Job Performance | Pearson Correlation Coefficient | 1 | -0.913 |
|  | Significance Level |  | .000 |
|  | Sample size | 247 | 247 |
| Technological Complexity | Pearson Correlation Coefficient | -0.913 | 1 |
|  | Significance Level | .000 |  |
|  | Sample size | 247 | 247 |

The results from the Pearson correlation test shown in Table 6 indicate that the significance level is zero and less than 0.05. This demonstrates a significant relationship between technology complexity and job performance, thereby confirming the third subsidiary hypothesis of the study. Additionally, the Pearson correlation coefficient is (-0.913), indicating a significant negative impact of technology complexity on job performance.

Pearson Correlation Test (Second Subsidiary Hypothesis): Table 7

|  |  | Job Performance | Technology Invasion |
|---|---|---|---|
| Job Performance | Pearson Correlation Coefficient | 1 | -0.838 |
|  | Significance Level |  | .000 |
|  | Sample size | 247 | 247 |
| Technology Invasion | Pearson Correlation Coefficient | -0.838 | 1 |
|  | Significance Level | .000 |  |
|  | Sample size | 247 | 247 |

The results from the Pearson correlation test in Table 7 show that the significance level is zero and less than 0.05. This indicates a significant relationship between technology invasion and job performance, thus confirming the fourth subsidiary hypothesis of the research. Moreover, the Pearson correlation coefficient is (-0.838), which points to a significant negative effect of technology invasion on job performance.

Pearson Correlation Test (Second Subsidiary Hypothesis): Table 8

|  |  | Job Performance | Technology-Induced Overload |
|---|---|---|---|
| Job Performance | Pearson Correlation Coefficient | 1 | -0.882 |
|  | Significance Level |  | .000 |
|  | Sample size | 247 | 247 |
| Technology-Induced Overload | Pearson Correlation Coefficient | -0.882 | 1 |
|  | Significance Level | .000 |  |
|  | Sample size | 247 | 247 |

The results from the Pearson correlation test in Table 8 indicate that the significance level is zero and less than 0.05. This demonstrates a significant relationship between technology-induced overload and job performance, thereby confirming the fifth subsidiary hypothesis of the research. Additionally, the Pearson correlation coefficient is (-0.882), indicating a significant negative effect of technology-induced overload on job performance. The results of the hypothesis testing of the study are presented in Table 9.

| Hypothesis Type | Pearson Coefficient | Confirmation/Rejection | Type of Relationship |
|---|---|---|---|
| Main | -0.940 | Confirmed | Negative |
| Subsidiary 1 | -0.860 | Confirmed | Negative |
| Subsidiary 2 | -0.890 | Confirmed | Negative |
| Subsidiary 3 | -0.913 | Confirmed | Negative |
| Subsidiary 4 | -0.838 | Confirmed | Negative |
| Subsidiary 5 | -0.882 | Confirmed | Negative |

**Discussion and Conclusion**

This research aims to analyze the relationship between technostress and job performance in the engineering and consulting company Tus Ab. The results from the hypothesis testing show that technostress, technology invasion, technology-induced overload, technology complexity, technology-induced insecurity, and technology uncertainty negatively correlate with job performance. This aligns with Kamour et al.'s (2013) research, which clearly demonstrated that technostress negatively correlates with organizational commitment and job satisfaction.

Many studies, such as those by Brenda Mack et al. (2010) and Mayer et al. (2002), have acknowledged the negative impact of technostress on organizational commitment. According to Daneshmandi et al. (2023), organizations must prioritize strategies that not only enhance job satisfaction through conducive work environments and recognition but also tackle technostress head-on.

Therefore, it is imperative for organizations, especially those heavily involved with technology, to provide continuous technical support and assistance to employees in using new technologies during times of crises and significant challenges. Technical support requires specific features: it must be accessible to employees, ready to respond during working hours, and comprised of knowledgeable and experienced individuals. To reduce the volume of work information and its less essential aspects, the entry of unnecessary data and emails should be limited, and all information should be organized using the most appropriate information systems, utilizing knowledge-based companies for information management. Managers should inform employees about approved programs before implementation, seek their opinions before using new systems, and ask for their help in adapting the new system to meet the organization's needs and objectives, as well as modifying the system to meet employees' needs.

On the other hand, among the dimensions of technostress, technology complexity has the most significant negative relationship with job performance, while technology invasion has the least. Taking into account this finding, it has been demonstrated that measures focused on improving affective and normative commitment can be especially beneficial in alleviating the negative psychological impacts linked with the pervasive presence of technology (Bai & Vahedian, 2023).

Table 10 provides practical suggestions for each dimension of technostress to control its negative effects.

Recommendations: Table 10

| Section | Recommendations | Outcome |
|---|---|---|
| Training | - Training for employees who need to learn how to use new technologies.<br><br>- Encourage employees to share knowledge about working with new technologies.<br><br>- Promote teamwork and solving new problems through group collaboration. | - Facilitates learning<br><br>- Reduces employee errors<br><br>- Positively impacts productivity reduced by technostress.<br><br>- Reduces technology complexity<br><br>- Reduces technology uncertainty. |
| Support | - Provide continuous technical support and assistance to employees for using new technology and during times of crisis and significant challenges. The technical support system must be accessible to employees, ready to respond at any required time during working hours, and consist of knowledgeable and experienced individuals. | - Increases employee satisfaction and prevents work disruptions.<br><br>- Reduces the feeling of technology uncertainty. |
| Learning and Experimentation | - Create an interactive and open environment for employees and interaction with managers.<br><br>- Reward employees seeking to gain experience and learn new skills.<br><br>- Encourage employees to take risks and gain new experiences.<br><br>- Assist with new ideas for easy implementation and execution | - Reduces technology-induced insecurity. |
| Facilitating Technological Participation | - Inform employees about newly approved application programs.<br><br>- Before using new systems, seek employees' opinions and ask for their help in adapting the new system to meet needs and organizational goals, and | - Increases effectiveness and satisfaction.<br><br>- Enhances organizational commitment. |

|  | change the system according to employees' needs. |  |
|---|---|---|
| Limitation | - Limit the entry of unnecessary data and emails.<br><br>- Organize information and use information systems.<br><br>- Utilize knowledge-based companies for information management. | - Reduces technology-induced overload<br><br>- Reduces technology invasion |